\documentclass[a4paper]{jpconf}
\usepackage{graphicx}
\usepackage{slashed}
\usepackage{amsmath}
\usepackage{amsfonts}
\usepackage{amssymb}
\usepackage{wrapfig}
\usepackage{hyperref}
\usepackage{color}
\usepackage{subfig}
\usepackage{multirow}

\begin{document}


\title{Phenomenology of additional scalar bosons at the LHC}

\author{Mukesh Kumar$^{a, 1}$,
  Stefan von Buddenbrock$^{b, 2}$, Nabarun Chakrabarty$^{c, 3}$,
  Alan S. Cornell$^{a, 4}$, Deepak Kar$^{b,5}$, Tanumoy Mandal$^{d,6}$, Bruce Mellado$^{b,7}$,
  Biswarup Mukhopadhyaya$^{c,8}$, Robert G. Reed$^{b,9}$ and Xifeng Ruan$^{b,10}$}

\address{$^{a}$ National Institute for Theoretical Physics; School of Physics and Mandelstam Institute for Theoretical Physics, University of the Witwatersrand, Johannesburg, Wits 2050, South Africa.}
\address{$^{b}$ School of Physics, University of the Witwatersrand, Johannesburg, Wits 2050, South Africa.}
\address{$^{c}$ Regional Centre for Accelerator-based Particle Physics, Harish-Chandra Research Institute, Chhatnag Road, Jhusi, Allahabad - 211 019, India.}
\address{$^{d}$ Department of Physics and Astronomy, Uppsala University, Box 516, SE-751 20 Uppsala, Sweden.}
\ead{$^{1}$mukesh.kumar@cern.ch,
$^{2}$stef.von.b@cern.ch,
$^{3}$nabarunc@hri.res.in,
$^{4}$alan.cornell@wits.ac.za,
$^{5}$deepak.kar@cern.ch,
$^{6}$tanumoy.mandal@physics.uu.se,
$^{7}$bruce.mellado@wits.ac.za,
$^{8}$biswarup@hri.res.in,
$^{9}$robert.reed@cern.ch,
$^{10}$xifeng.ruan@cern.ch}

\begin{abstract}
The confirmation of the Higgs boson in Run I data at the Large Hadron Collider (LHC) and the excesses in recent 
Run II data suggest scenarios beyond the Standard Model (SM). We pursue a study in a minimal model which is 
an extension of a scalar doublet in the SM known as two-Higgs doublet model (THDM). 
Following earlier suggestions two real scalars $\chi$ and $S$ have been introduced in the THDM where $\chi$ is 
treated as a candidate for dark matter. $\chi$ does not receive any vacuum expectation value ($vev$) in the model 
whereas the Higgs-like scalar $S$ acquires $vev$. This allows small mixing between the $CP$-even scalars of the 
THDM, $h$, $H$ and $S$. In this study the mass spectrum of new scalars is taken to be $2 m_h < m_H < 2 m_t$, 
$m_\chi < m_h/2$, $m_h \lesssim m_S \lesssim m_H - m_h$, $m_A > 2 m_t$ and $m_H^\pm < m_A$,
where $m_h$ and $m_t$ is masses of the SM Higgs and top-quark respectively, $m_H, m_A$ and $m_{H^\pm}$ are 
the masses of the heavy $CP$-even scalar $H$, $CP$-odd scalar $A$, and charged Higgs $H^\pm$, respectively. 
A partial list of potential search channels at the LHC has been provided with possible phenomenological 
consequences. The expected phenomenology and constraints on parameters are also discussed in a model-independent 
approach .  

\end{abstract}

\section{Introduction}
After the confirmation of the Higgs boson ($h$) in Run I data at the Large Hadron Collider (LHC)~\cite{Englert:1964et,Higgs:1964ia,Higgs:1964pj,Guralnik:1964eu,ATLAS:2012gk,CMS:2012gu}, it is pragmatic to move forward in 
searches of physics beyond the Standard Model (BSM). In this context we already have well-defined models with rich particle 
spectrum which shall be supplemented through observed excesses in the experimental data. 
Some of the excesses in Run I data at the ATLAS and CMS, namely, the excess in transverse momentum ($p_T$) spectrum of the
Higgs boson in $h\to \gamma\gamma$ and $h\to ZZ\to 4l$~\cite{Aad:2014lwa, Aad:2014tca, Khachatryan:2015rxa, CMS:2015hja}
channel has been addressed in Ref.~\cite{vonBuddenbrock:2015ema} by corroborating
other source of excesses in heavy scalar searches through di-Higgs boson production~\cite{Aad:2015xja, CMS:2014ipa, Khachatryan:2015tha, Khachatryan:2014jya} and in associated production of top-quarks~\cite{Aad:2014lma, Aad:2015iha, Aad:2015gra, Khachatryan:2014qaa}.
Recent excess in di-photon channel around 750~GeV~\cite{ATLAS:750,CMS:2016owr} in early data of Run II at the LHC also
provides hints of new physics BSM where it could be assumed to arise through a scalar or pseudo-scalar resonance around this
mass.

These excesses can be explained by introducing a particle spectrum BSM in an effective theory as well as in a proper model
by considering the appropriate sources of constraints in the theory and known limits on parameter space in that particular model.
It is evident that this new particle spectrum comes with new phenomenology and helps in not only explaining the observed data
but also in searches beyond a particular model(s). In this proceeding we intent to address the phenomenology of scalars
BSM, assuming the particle spectrum of two-Higgs doublet model (THDM) in addition with other real 
singlets $\chi$ and $S$ introduced in Refs.~\cite{Kumar:2016vut, vonBuddenbrock:2016rmr}. 

\section{The Model}
Before moving towards the phenomenological implications of the additional scalars BSM, we first discuss the particle spectrum and 
parameters introduced in a model dependent as well as in a model independent approach. 
A simple extension of the SM is known as the THDM with one more additional scalar doublet, where this additional scalar doublet 
transform under the SM gauge symmetries similar to the Higgs field in the SM. After spontaneous electroweak symmetry breaking,
five physical Higgs particles are left in the spectrum of THDM, two $CP$-even states, $H$ (heaviest) and $h$ (lightest), one $CP$-odd scalar, $A$, and one charged Higgs pair, $H^\pm$. The ratio of the two vacuums ($v_1, v_2$) of corresponding doublets is a
free parameter of the theory, $\tan\beta = v_2/v_1$ in addition with the masses of the scalars\footnote{It is to be noted that the two
$CP$-even scalars $h$ and $H$ may have different interpretations in terms of masses in comparison to the SM Higgs boson. In our
case we assume the lighter $h$ to aligned with the SM Higgs boson.} and the mixing angle $\alpha$ between $h$ and $H$. Based on
different choices of symmetries and couplings to fermions different types of THDM is available, $viz.$, Type-I, Type-II, Lepton-specific
or Flipped THDMs. For more details of this model we refer the Refs.~\cite{gunion, Branco:2011iw}.

In a model-independent approach i.e. $via$ an effective theory the authors of Refs.~\cite{vonBuddenbrock:2015ema, Kumar:2016vut, vonBuddenbrock:2016rmr} introduced three real scalars $H,$\footnote{$H$ may not be same as THDM heavier $CP$-even scalar. The difference is also noticed since in this approach it is considered to be a real singlet scalar while in THDM it is a part of doublet which appear after spontaneous electroweak symmetry breaking.} $\chi$ - introduced as a dark matter (DM) candidate whose
signature is a source of missing energy in the colliders, and $S$ - a Higgs-like real singlet scalar. In these studies, it is found
necessary to fit all data that the heavier $H$ should have a large branching ratio in the channel $H \to h\chi\chi$, where $h$ is the 
Standard Model (SM) Higgs boson. However, this explanations can also be accommodated by introducing the on-shell participation 
of $S$ in the decay of $H \to S h, S \to \chi\chi$. To achieve the fittings, the free parameters of the theory are fixed by considering
various observed limits in $VV$ ($V = W^\pm, Z$) resonance searches~\cite{Aad:2015agg, Aad:2015kna, Khachatryan:2015cwa} 
and the DM constraints.  

As an extension to this model-independent approach, it is interesting to add $\chi$ and $S$ in a THDM as a model-dependent 
scenarios. In that case all new parameters of interactions among these scalars follow the constraints from THDM with few additional
parameters in the theory. The details of this model can be found in Ref.~\cite{vonBuddenbrock:2016rmr}, where $\chi$ does not
acquire vacuum expectation value ($vev$) while $S$ acquires $vev$ in result there is small mixing between $h, H$ and $S$ is
considered to obtain $hHS$ coupling which should respect the bounds from the Higgs data. 
        
\section{Phenomenology}
In this section we discuss the phenomenology associated with the scalars introduced in previous sections which is useful in 
explaining the observed data in experiments as well as to provide hints for searches at particular energy and parameter choices in the
colliders like the LHC. The observed final states are highly dependent on the choice(s) of masses of these scalars after production 
rates $via$ gluons, quarks or gluons and quarks.  

Following the results of Ref.~\cite{vonBuddenbrock:2015ema}, we consider following mass ranges for THDM scalars with additional
real scalars $\chi$ and $S$ to explain the phenomenology:
\begin{itemize}
\item [(a)] Light Higgs: $m_h = 125$~GeV (\text{assuming as the SM Higgs}),
\item [(b)] Heavy Higgs: $2 m_h < m_H < 2 m_t$,
\item [(c)] $CP$-odd Higgs: $m_A > \left(m_H + m_V\right)$, where $\left( V = W^\pm, Z \right)$,
\item [(d)] Charged Higgs: $\left(m_H + m_V\right) < m_{H^\pm} < m_A$,
\item [(e)] Additional scalars $\chi$, $S$: $m_\chi < m_h/2$ and $m_h \lesssim m_S \lesssim \left(m_H - m_h\right)$.
\end{itemize}
Based on these masses we can study the branching fractions of THDM scalars into the SM particles and additional scalars
$\chi$ and $S$ as listed in Table~\ref{tab:i} with following production channels:
\begin{itemize}
\item [(a)] $g g \to h$, $H$, $A$, $S$,
\item [(b)] $p p \to t H^- (\bar t H^+)$, $t H^-\bar b + \bar t H^+ b$, $H^+ H^-$, $H^\pm W^\pm$.
\end{itemize}
A partial list of interesting searches are summarised in Table~\ref{tbl:los} which leads striking signatures of clean final states
involving the scalars introduced in this proceedings. Mostly we discussed leptonic final states with missing energy signatures
through the decays of heavy scalars. The same-sign leptonic channels provides prominent signatures for any new physics
models beyond the SM. Apart from these search list 
\begin{itemize}
\item[(1)] It is interesting to constrain the parameter space of THDM and associated parameters with $\chi$ and $S$ by 
explaining the distortion in the $p_T$ spectrum of the Higgs boson in $h \to \gamma\gamma$ and $h \to ZZ \to 4l$ channels as
in Ref.~\cite{vonBuddenbrock:2015ema}. Also the consequences of introducing $S$ to explain the large branchings in the effective
theory due to $H\to h\chi\chi$ can be understood in both model independent and dependent approaches by allowing $H \to h S$ 
and $S \to \chi\chi$.   
\item[(2)] The first LHC data at $\sqrt{s} = 13$~TeV observed an excess in $\gamma\gamma$ final state peaked at the invariant
mass around 750~GeV~\cite{ATLAS:750,CMS:2016owr} with the best fit width of the resonance as $~45$~GeV. Here according to
our parameter choice that resonance could be a $CP$-odd scalar $A$. And further based on the choices of mass and width of
this resonance the model parameters can be constrained.   
\end{itemize}
Few analysis of interesting final states with selected leptonic signatures are discussed in the Ref.~\cite{vonBuddenbrock:2016rmr}
by considering $H \to 4 W \to 4l$ with missing energy and associated $H$ production with top quarks and decays as 
$t(t)H \to 6 W \to l^\pm l^\pm l^\pm +X $. With this note we summarise our work in the next section.
\begin{table*}[tbp]
\renewcommand{\arraystretch}{1.15}
\centering
\begin{tabular}{c|c|l}
\hline
 \textbf{S. No.}                 & \textbf{Scalars}       & \textbf{Decay modes} \\ \hline
\texttt{D.1} & $h$ & $b\bar b$, $\tau^+ \tau^-$, $\mu^+\mu^-$, $s\bar s$, $c\bar c$, $gg$, $\gamma\gamma$, $Z\gamma$,
  $W^+W^-$, $ZZ$     \\
\texttt{D.2} & $H$ & \texttt{D.1}, $hh$, $SS$, $Sh$ \\ 
\texttt{D.3} & $A$ & \texttt{D.1}, $t\bar t$, $Zh$, $ZH$, $ZS$, $W^\pm H^\mp$      \\
\texttt{D.4} & $H^\pm$ & $W^\pm h$, $W^\pm H$, $W^\pm S$      \\
\texttt{D.5} & $S$ & \texttt{D.1}, $\chi\chi$      \\
\hline
\end{tabular}
\caption{\label{tab:i} List of possible decay modes of 2HDM scalars to the SM particles and additional scalars
based on the mass choices of all scalars as described in the text. Note that we are not interested in $h\to \chi\chi$
decay, instead we prefer $S\to \chi\chi$ decay mode. In the final states of all decay we consider $W^\pm$, $Z \to$~leptonic
decay modes.}
\end{table*}

\begin{table*}[tbp]
        \renewcommand{\arraystretch}{1.15}
        \centering
        \begin{tabular}{c|l|l} \hline
                \textbf{Scalar} & \textbf{Production mode} & \textbf{Search channels} \\
                \hline
                \multirow{9}{7pt}{$H$} & $gg\to H, Hjj$ ($gg$F and VBF) & Direct SM decays as in \autoref{tab:i} \\
                & & $\to SS/Sh\to 4W\to 4\ell$ + MET \\
                & & $\to hh\to \gamma\gamma b\bar{b},~b\bar{b}\tau\tau,~4b,~\gamma\gamma WW$ etc. \\
                & & $\to Sh$ where $S\to\chi\chi\implies \gamma\gamma,~b\bar{b},~4\ell$ + MET \\
                \cline{2-3}
                & $p p \to Z(W^\pm) H~(H\to SS/Sh)$ & $\to$ $6(5) l$ + MET \\
                & & $\to 4(3) l + 2j$ + MET \\
                & & $\to 2(1) l + 4j$ + MET \\
                \cline{2-3}
                & $p p \to t \bar t H, (t + \bar t)H~(H \to S S / Sh)$ & $\to2 W + 2 Z$ + MET and $b$-jets \\
                & & $\to6W\to3~\text{same sign leptons}$ + jets and MET \\
                \hline
                \multirow{4}{7pt}{$H^\pm$} & $p p \to t H^\pm~(H^\pm\to W^\pm H)$ & $\to6W\to3~\text{same sign leptons}$ + jets and MET \\
                \cline{2-3}
                & $p p \to t b H^\pm~(H^\pm\to W^\pm H)$ & Same as above with extra $b$-jet \\
                \cline{2-3}
                & $pp\to H^\pm H^\mp~(H^\pm\to HW^\pm)$ & $\to6W\to3~\text{same sign leptons}$ + jets and MET \\
                \cline{2-3}
                & $p p \to H^\pm W^\pm~(H^\pm\to HW^\pm)$ & $\to6W\to3~\text{same sign leptons}$ + jets and MET \\
                \hline
                \multirow{5}{7pt}{$A$} & $gg\to A$ ($gg$F) & $\to t\bar{t}$ \\
                & & $\to\gamma\gamma$ \\
                \cline{2-3}
                & $gg\to A\to Z H~(H\to SS/Sh)$ & Same as $pp\to ZH$ above, but with resonance \\&& structure over final state objects \\
                \cline{2-3}
                & $gg\to A\to W^\pm H^\mp (H^\mp\to W^\mp H)$ & $6W$ signature with resonance \\&& structure over final state objects \\
                \hline
        \end{tabular}
        \caption{A list of potential search channels arising from the addition of the new scalars presented in this proceedings. This list is by no means complete, but contains clean search channels (mostly leptonic) which could make for striking signatures in the LHC physics regime. Note that in the mass ranges we are considering, $H$ almost always decays to $SS$ or $Sh$, where $S$ and $h$ are likely to decay to $W$s or $b$-jets.}
        \label{tbl:los}
\end{table*} 

\section{Summary}
In this proceedings we discuss the phenomenology of the scalars beyond the Standard Model. 
These scalars are introduced in a model independent way as an effective theory and also in a model dependent approach
- considering the two-Higgs doublet model. 
A logical way of introducing $\chi$ and $S$ has been discussed. A list of search strategies has been drawn
and based on recent observed excesses in ATLAS and CMS data, a sketch has been suggested incorporating the mass spectrum
of the introduced scalars. In most of the cases we considered the leptons with missing energy final states with a significant number
of same-sign leptons make the searches to be clean and striking.

Here it is also important to summarise the present constraints on the two-Higgs doublet model parameters from Run-I data at the
LHC.
In Ref.~\cite{cms-16-007} a summary of exclusion limits were presented for several searches on additional heavy Higgs bosons in physics BSM. In the case of THDM, two benchmark scenarios Type-I and Type-II were considered. The Type-I (II) THDM is
generally constrained to $\cos\left(\beta - \alpha \right) \lesssim 0.5 (0.2)$, $m_H \lesssim 380 (\approx 380)$ and 
$\tan\beta \lesssim 2$ (all) of the heavy Higgs boson. These constraints have been obtained by considering the decay channels
$A/H/h \to \tau\tau$, $H\to WW/ZZ$, $A\to ZH(llbb)$ and $A\to ZH(ll\tau\tau)$. We hope that our work towards this direction in Run-II 
data will show interesting features.
      

\end{document}